\documentclass[11pt]{article}
\usepackage{fullpage}
\usepackage[utf8]{inputenc}
\usepackage[english]{babel}
\usepackage[T1]{fontenc}
\usepackage{amsmath,amssymb,amsfonts,amsthm}
\usepackage{url}
\usepackage{hyperref}
\usepackage{cleveref}
\usepackage{graphicx}
\usepackage[ruled,vlined, linesnumbered]{algorithm2e}

\newtheorem*{theorem*}{Theorem}
\newtheorem{lemma}{Lemma}
\newtheorem{prop}{Proposition}

\newtheorem{cor}{Corollary}

\usepackage{authblk}

\newcommand{\Z}{\mathbb{Z}}

\newcommand{\ket}[1]{|{#1}\rangle}

\newcommand{\Prob}{\mathbb{P}}
\newcommand{\mat}{{\bf M}}

\newcommand{\PauliWithPhase}{{\cal P}}
\newcommand{\Pauli}{{\overline{\cal P}}}

\DeclareMathOperator{\fail}{fail}
\DeclareMathOperator{\sample}{sample}

\newcommand{\circuit}{{\cal C}}

\DeclareMathOperator{\eff}{eff}

\newcommand{\propagation}[1]{\protect\overrightarrow{#1}}
\newcommand{\backpropagation}[1]{\protect\overleftarrow{#1}}

\usepackage{xcolor}

\title{Simulation of noisy Clifford circuits without fault propagation}

\author[]{Nicolas Delfosse, Adam Paetznick}

\affil[]{Microsoft Quantum, Redmond, Washington 98052, USA}

\begin{document}

\maketitle

\date

\begin{abstract}
The design and optimization of a large-scale fault-tolerant quantum computer architecture relies extensively on numerical simulations to assess the performance of each component of the architecture.
The simulation of fault-tolerant gadgets, which are typically implemented by Clifford circuits, is done by sampling circuit faults and propagating them through the circuit to check that they do not corrupt the logical data.
One may have to repeat this fault propagation trillions of times to extract an accurate estimate of the performance of a fault-tolerant gadget.
For some specific circuits, such as the standard syndrome extraction circuit for surface codes, we can exploit the natural graph structure of the set of faults to perform a simulation without fault propagation.
We propose a simulation algorithm for all Clifford circuits that does not require fault propagation and instead exploits the mathematical structure of the spacetime code of the circuit.
Our algorithm, which we name adjoint-based code (ABC) simulation, relies on the fact that propagation forward is the adjoint of propagation backward in the sense of Proposition~3 from~\cite{delfosse2023spacetime}.
We use this result to replace the propagation of trillions of fault-configurations by the backward propagation of a small number of Pauli operators which can be precomputed once and for all.
\end{abstract}

At the core of the architectures of fault-tolerant quantum computers are quantum error correction codes such as surface codes~\cite{dennis2002topological, raussendorf2007fault, fowler2012surface}, Floquet codes~\cite{hastings2021dynamically, paetznick2023performance, gidney2022benchmarking} or quantum LDPC codes~\cite{tillich2013quantum, breuckmann2021quantum, panteleev2022asymptotically, leverrier2022quantum, tremblay2022constant}.
Universal fault-tolerant quantum computing requires defining a set of logical operations by way of (physical) quantum circuits. This may, for instance, include idle gates, lattice surgery~\cite{horsman2012surface}, magic state distillation and state injection circuits~\cite{bravyi2005universal}. 
Characterizing and optimizing all these circuits, sometimes called gadgets, for a given specification of qubits, gate set, connectivity and noise model requires substantial numerical simulation.

\begin{figure}
    \centering
    \includegraphics[scale=.5]{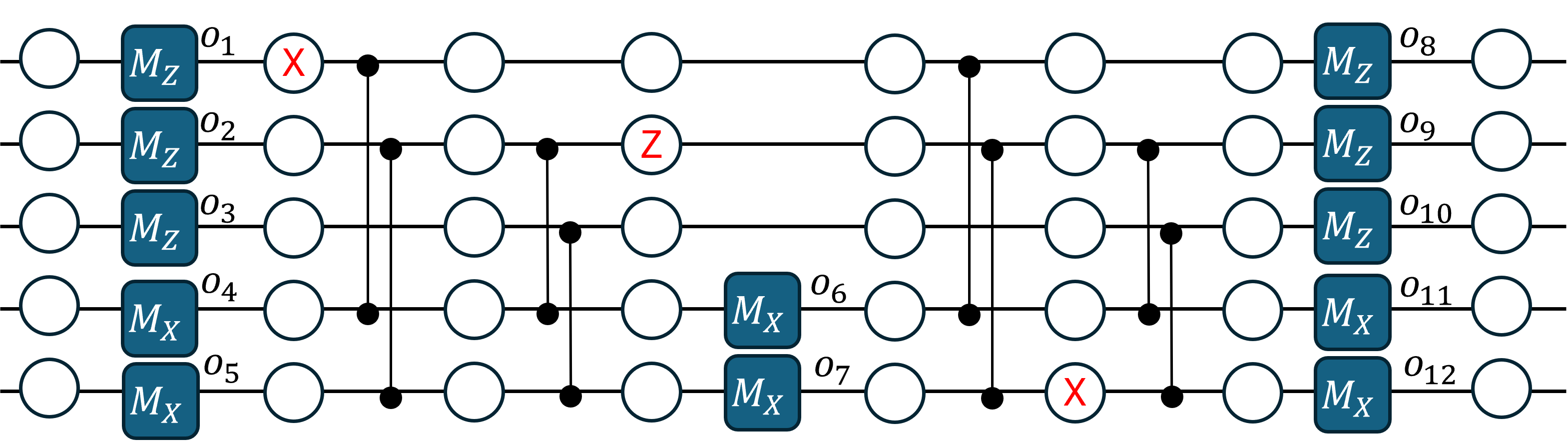}
    
    (a)
    
    \includegraphics[scale=.5]{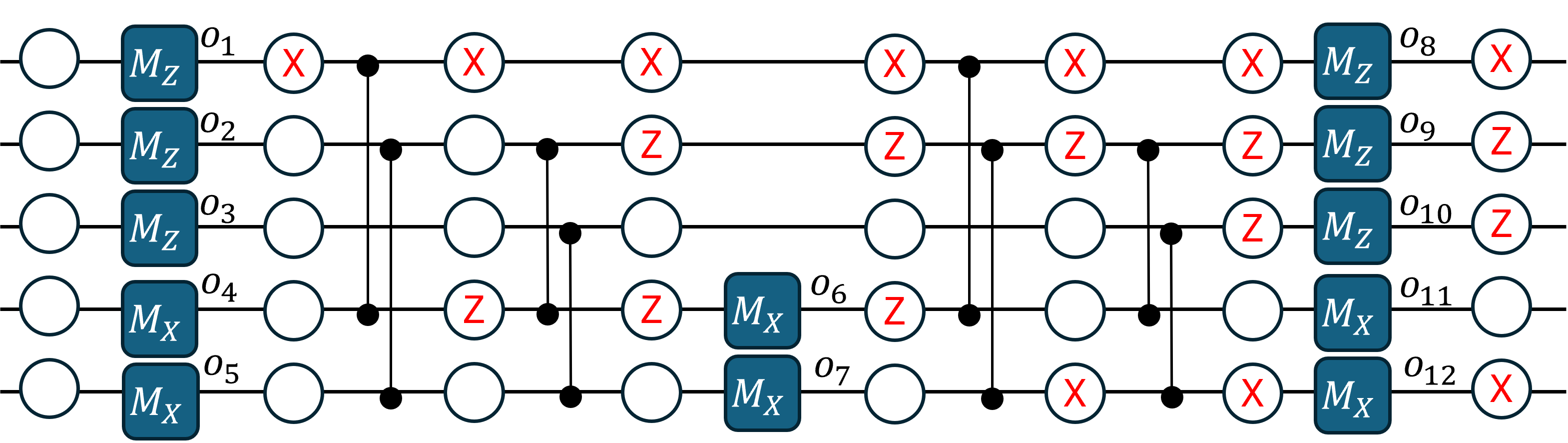}
    
    (b)

    \caption{A Clifford circuit made with Pauli measurements and CZ gates. This circuit implements the measurement of the stabilizer generators $Z_1 Z_2$ and $Z_2Z_3$ of the repetition code (repeated twice) on the top three qubits using the two bottom qubits as ancillas.
    The white circles indicate the locations of potential faults.
    In the absence of faults, the outcome bit-string $o \in \Z_2^{12}$ satisfies the following checks:
    (i) $o_1 + o_2 + o_4 + o_6 = 0$,
    (ii) $o_2 + o_3 + o_5 + o_7 = 0$,
    (iii) $o_1 + o_2 + o_6 + o_{11} = 0$,
    (iv) $o_2 + o_3 + o_7 + o_{12} = 0$.
    If any of these checks is violated, we know that a fault must have occurred in the circuit. This allows us to detect or correct some circuit faults.
    (a) A fault configuration represented by a Pauli operator $F$, called a fault operator. We can think of $F$ as a Pauli operator acting on qubits placed on the spacetime locations of the circuit (white circles). 
    (b) The cumulant $\propagation{F}$ is obtained by propagating the faults of $F$ through the circuit. The propagation through measurements is trivial and the propagation through a unitary gate is obtained by conjugating the input faults by the gate.
    By inspecting the cumulant $\propagation{F}$, one can determine whether $F$ flips the outcomes $o_i$ and then compute the checks. Indeed, the outcome $o_i$ is flipped iff $\propagation{F}$ anti-commutes with the measured operator $o_i$ at the time step right before the measurement.
    In this example, $F$ flips the outcomes $o_6$ and $o_8$.
    The syndrome, that is the value of the four checks (i), (ii), (iii), (iv), is $(1, 0, 1, 0)$.
    }
    \label{fig:standard_simulation}
\end{figure}

A typical scenario is that you have a Clifford circuit implementing a fault-tolerant gadget and you want to estimate the failure rate of this gadget for different noise parameters.
We consider the standard circuit-noise model~\cite{dennis2002topological}. Each circuit operation is followed by a random Pauli error acting on its support and measurement outcomes are flipped with some probability.
The standard approach to estimating the performance of a Clifford circuit proceeds with the following steps.
\begin{enumerate}
    \item Sample circuit faults according to the noise model.
    \item Propagate these faults through the circuit using the Gottesman-Knill algorithm~\cite{gottesman1998heisenberg} to determine their effect on the measurement outcomes and the output qubits.
    \item Run some classical post-processing based on the measurement outcomes flipped by the faults.
    \item Determine if the faults lead to a failure of the gadget.
\end{enumerate}
The classical post-processing may include computation of syndrome data, execution of a decoder, or computing parities of measurement outcomes based on which the gadget performs post-selection.
Figure~\ref{fig:standard_simulation} shows an example of computation of the syndrome using this approach.
This simulation is typically repeated a large number of times to generate enough data to obtain a good estimate of the failure rate of the gadget. 
Say, for example, that we want to probe a three-parameter noise model. 
If we select only ten values for each noise parameter and if for each triple of values we need a billion samples to reach a sufficiently small error bar for the corresponding data point, this results in one trillion repetitions of the previous steps.

The dominant cost of this simulation is generally running the decoder and propagating Pauli faults through the circuits.
Low-complexity decoders have been designed~\cite{fowler2012surface, delfosse2021almost} and fast decoder implementations are available~\cite{higgott2023sparse, wu2023fusion}.
In what follows, we propose a simulation protocol that provides the same estimate of the failure rate of a Clifford gadget with circuit-noise without any fault propagation.

\begin{figure}
    \centering
    \includegraphics[scale=.5]{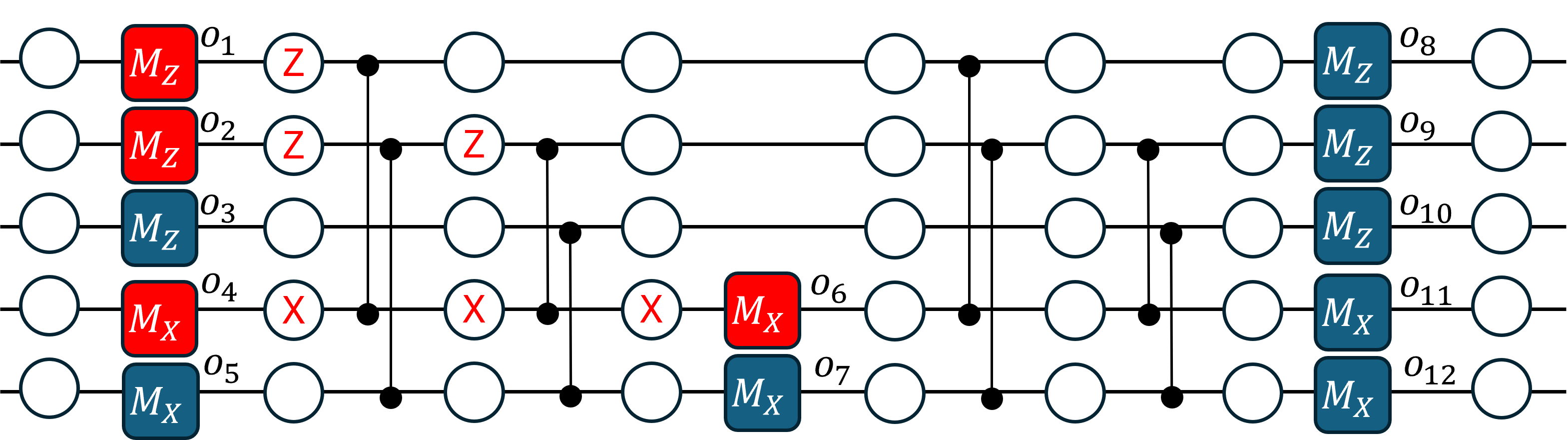}
    
    (a) Backpropagation of the check $o_1 + o_2 + o_4 + o_6 = 0$.
        
    \includegraphics[scale=.5]{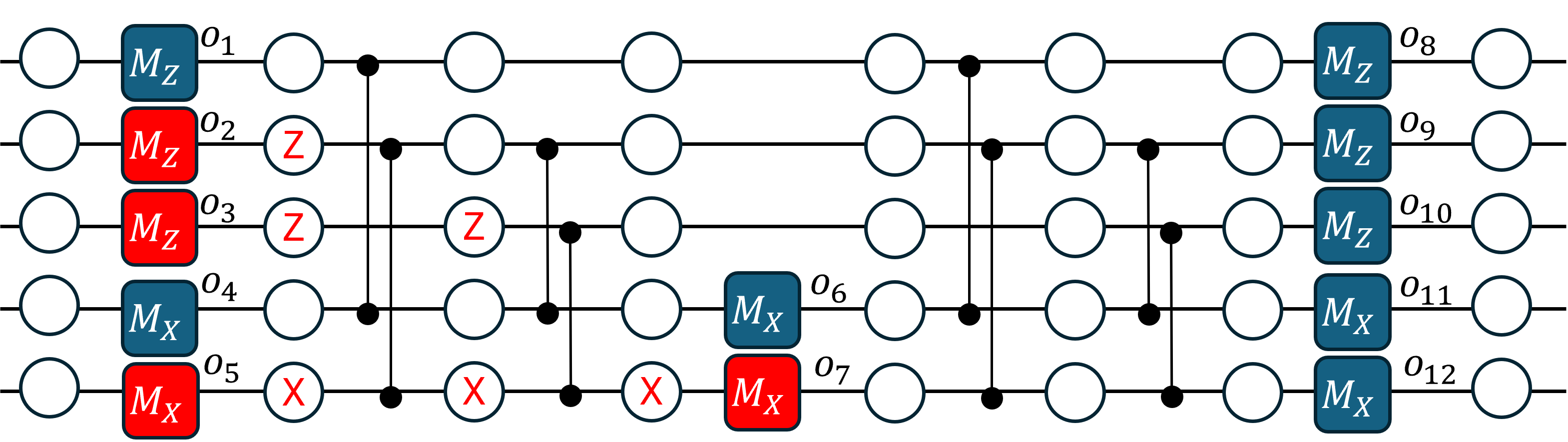}
    
    (b) Backpropagation of the check $o_2 + o_3 + o_5 + o_7 = 0$.
        
    \includegraphics[scale=.5]{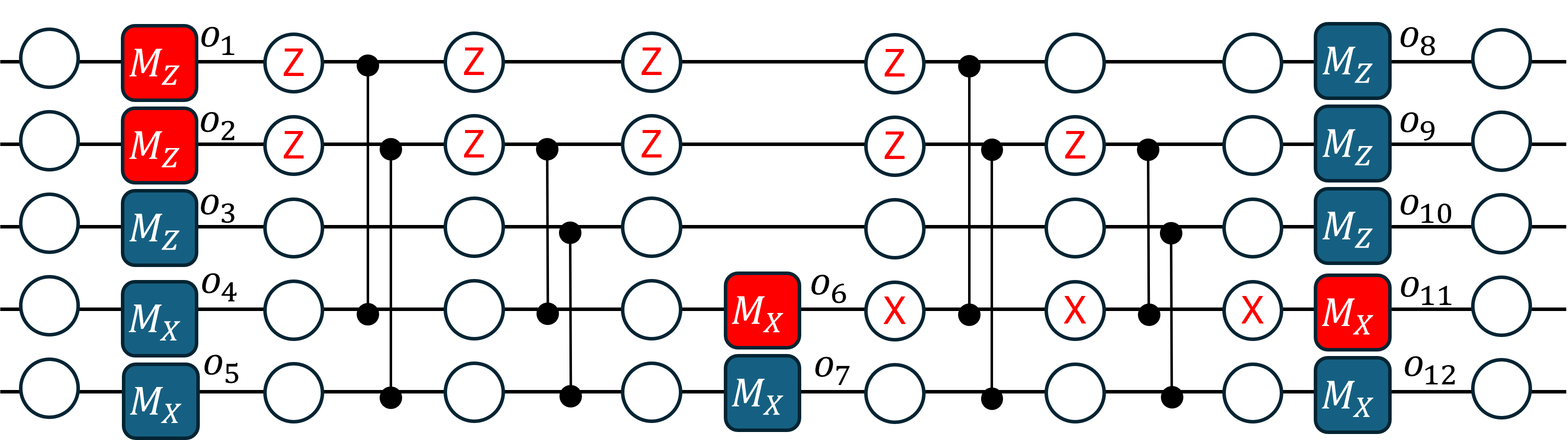}
    
    (c) Backpropagation of the check $o_1 + o_2 + o_6 + o_{11} = 0$.
        
    \includegraphics[scale=.5]{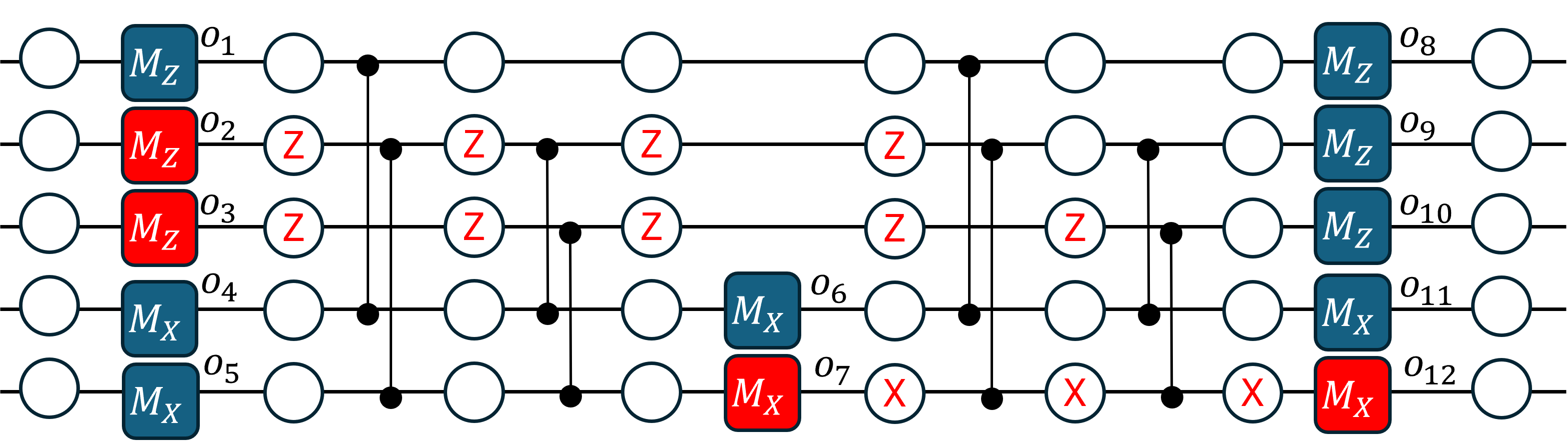}
    
    (d) Backpropagation of the check $o_2 + o_3 + o_7 + o_{12} = 0$.
        
    \caption{The standard approach to compute the syndrome of a set of faults is through fault-propagation as shown in Figure~\ref{fig:standard_simulation}.
    This figure illustrates our ABC simulation scheme that does not require fault-propagation. The syndrome is computed using the backpropagation of the checks.
    The backpropagation of the checks (i), (ii), (iii) and (iv) of Figure~\ref{fig:standard_simulation} is represented in (a), (b), (c) and (d).
    They are obtained by placing the measured operators (involved in the check) right before they are measured and by backpropagating them~\cite{delfosse2023spacetime}.
    If $\backpropagation{G}$ denotes the backpropagated operator corresponding to a check, then value of the check is non-trivial iff $F$ anti-commutes with $\backpropagation{G}$. By inspecting the commutation between $F$ and the operators (a), (b), (c), (d), we recover the syndrome $(1, 0, 1, 0)$ of $F$ without needing $\propagation{F}$.
    }
    \label{fig:simulation_without_propagation}
\end{figure}

The basic idea of our ABC simulation to leverage the spacetime code structure~\cite{delfosse2023spacetime}; see also~\cite{bacon2015sparse, gottesman2022spacetimecode}.
More specifically, Proposition~3 from~\cite{delfosse2023spacetime} (replicated below in Proposition~\ref{prop:key_prop_adjoint}) allows us to replace the (forward) propagation of Pauli faults through a Clifford circuit by the backward propagation of stabilizer generators of the spacetime code.
As a result, instead of propagating faults trillions of times, we only need to precompute the backward propagation of the spacetime generators once.
Similarly, we precompute the backward propagation of some logical operators to determine if the protocol fails.
Figure~\ref{fig:simulation_without_propagation} illustrates our propagation-free simulation method.

The paper is organized as follows.
Our notations and assumptions are described in Section~\ref{sec:clifford_circuits} and we briefly review the correction of circuit faults based on the outcome code~\cite{delfosse2023spacetime} in Section~\ref{sec:outcome_code}.
The standard simulation protocol is reviewed in Section~\ref{sec:standard_simulation} and our ABC simulation protocol without fault propagation is presented in Section~\ref{sec:sim_wo_prop}.
Finally, we discuss the application of the ABC simulation strategy to the simulation of large circuit in Section~\ref{sec:large_sim}.

\section{Noisy Clifford circuits} \label{sec:clifford_circuits}

We consider Clifford circuits made with unitary Clifford gates and measurements of Pauli operators. We follow the assumptions and notations of~\cite{delfosse2023spacetime}, which we briefly review now.

We consider a circuit acting on $n$ qubits with depth $\Delta$.
Denote by $\PauliWithPhase_n$ the set of $n$-qubit Pauli operators and by $\Pauli_n$ its quotient by the phase operators $\{\pm I, \pm i I\}$.
A configuration of faults in the circuit is represented by a {\em fault operator}, which is a Pauli operator $F \in \Pauli_{n(\Delta+1)}$ acting on $n (\Delta+1)$ qubits. We can think of a fault operator as a Pauli operator acting on qubits placed on half-integer times steps of the circuit and indexed by pairs $(\ell + 0.5, q)$ where $\ell \in \{0, 1, \dots, \Delta\}$ is a level of the circuit and $q \in \{1, \dots, n\}$ is a qubit.

A {\em circuit-noise model} for a circuit $\circuit$ is defined to be a probability distribution, denoted $\Prob_\circuit$ over the set of fault operators.
Such a noise model includes Pauli faults and measurement outcomes flips that can be represented by a Pauli error before and after a measurement.

We assume that the circuit contains $n_m$ measurements. Each run of the circuit produces an {\em outcome bit-string} $o \in \Z_2^{n_m}$ whose $i$th component is the outcome of the $i$th measurement.
Based on the outcome bit-string, a {\em logical outcome} $\ell \in \Z_2^{n_\ell}$ is computed by applying a binary matrix $\mat_\ell$ to $o$, that is $\ell^T = \mat_\ell o^T$ where $\mat_\ell \in M_{n_\ell, n_m}(\Z_2)$.
Each component of the vector $\ell$ is a logical bit which is obtained by taking the parity some measurement outcomes\footnote{These outcomes are indicated by the corresponding row of $\mat_\ell$.}.
For example, the outcome of a logical $X$ measurement in the surface code is obtained by measuring all the qubits of a logical patch in the $X$ basis and then taking the parity of the measurement outcomes along a line of qubits supporting a logical $X$ operator~\cite{fowler2018low}. In the presence of noise this logical outcome bit must be corrected by the decoder.

The {\em effect} of a fault operator $F$ is defined to be the pair $\eff(F) = (f, E)$ where $f \in \Z_2^{n_m}$ represents the measurement outcome flips induced by $F$ and $E \in \Pauli_{n}$ is the residual error on the qubits at the end of the circuit when $F$ occurs.
Recall that $f_j = 1$ iff $F$ leads to a flip of the outcome of the $j$th measurement of the circuit.
We use the notation $\eff_{m}(F) = f$ and $\eff_q(F) = E$ for the effect on measurement outcomes and the effect on qubits.

\section{Correction of circuit faults using the outcome code}
\label{sec:outcome_code}

In~\cite{delfosse2023spacetime}, we proved that the outcome bit-string belongs to a linear code (up to a relabelling of the measurement outcomes) that we call the {\em outcome code} and we explain how to correct circuit faults using this code. This leads to a general correction protocol including a broad class of fault tolerant gadgets for stabilizer codes~\cite{gottesman1997stabilizer}, surface codes~\cite{dennis2002topological, fowler2012surface}, color codes~\cite{bombin2006topological} or Floquet codes~\cite{hastings2021dynamically}.

The correction of circuit faults based on the outcome code works as follows.
After extracting the outcome bit-string $o$, compute its syndrome $s \in \Z_2^{n_s}$ by applying a binary matrix $\mat_s \in M_{n_s, n_o}(\Z_2)$ to $o$, that is $s^T = \mat_s o^T$.
The matrix $\mat_s$ can be efficiently generated using Algorithm~1 of~\cite{delfosse2023spacetime}.
In the absence of faults, the syndrome is trivial.
If the faults corresponding to a fault operator $F$ occur, we obtain the syndrome $s^T = \mat_s f^T$ where $f = \eff_m(F)$, which depends only on $F$.
Moreover, $F$ flips some logical outcomes. The indicator vector of the flipped logical bits is the vector $\bar f$ given by $\bar f^T = \mat_\ell f^T$.

The syndrome matrix corresponding to the four checks of Figure~\ref{fig:standard_simulation} and Figure~\ref{fig:simulation_without_propagation} is
\setcounter{MaxMatrixCols}{20}
\begin{align} \label{eq:syndrome_matrix}
    \mat_s =
    \begin{pmatrix}
    1 & 1 & 0 & 1 & 0 & 1 & 0 & 0 & 0 & 0 & 0 & 0 \\
    0 & 1 & 1 & 0 & 1 & 0 & 1 & 0 & 0 & 0 & 0 & 0 \\
    1 & 1 & 0 & 0 & 0 & 1 & 0 & 0 & 0 & 0 & 1 & 0 \\
    0 & 1 & 1 & 0 & 0 & 0 & 1 & 0 & 0 & 0 & 0 & 1
    \end{pmatrix} \cdot
\end{align}
Here, we have $n_m = 12$ and $n_s = 4$.
Each column of this matrix corresponds to an outcome bit and each row is the indicator vector of a check.
For example, the first row defines the check $o_1 + o_2 + o_4 + o_6 = 0$.
Using this circuit, one can measure a logical $Z$ operator for the repetition code. The logical outcome is simply $o_8$ and the corresponding logical matrix is
\begin{align} \label{eq:logical_matrix}
    \mat_\ell =
    \begin{pmatrix}
    0 & 0 & 0 & 0 & 0 & 0 & 0 & 1 & 0 & 0 & 0 & 0 \\
    \end{pmatrix} \cdot
\end{align}
Moreover, the effect of the fault $F$ from Figure~\ref{fig:standard_simulation} on measurement outcomes is 
\begin{align} \label{eq:effect}
    f =
    \begin{pmatrix}
    0 & 0 & 0 & 0 & 0 & 1 & 0 & 1 & 0 & 0 & 0 & 0 \\
    \end{pmatrix}
\end{align}
because $F$ induces a flip of the outcomes $o_6$ and $o_8$.
The syndrome of $F$ is $s = (1, 0, 1, 1)$ and the logical effect is $\bar f = (1)$.
Note that this example is intended for illustration purposes. The circuit is not fault tolerant.

A {\em decoder} $D$ is used to correct the logical outcome of the circuit.
It takes as an input the syndrome $s$ and returns a correction $D(s) = \bar f'$ to apply to the logical outcome $\ell$, {\em i.e.} replacing $\ell$ by $\ell + \bar f'$.
We say a {\em failure} occurs if the logical outcome $\ell + \bar f'$ after decoding is incorrect due to the presence of faults in the execution of circuit, that is iff $\bar f' \neq \bar f$.

Our goal is to design a Monte-Carlo simulation to estimate the {\em failure rate} $\Prob_\circuit(D(s) \neq \bar f)$ of the circuit.
We refer to such a simulation as a {\em circuit-noise simulation}.

For simplicity, we focus on failures to recover the logical outcome but our simulation protocol without fault propagation can be generalized to failures induced by residual errors on the output qubits of the circuit.
In that case, the decoder may also apply a correction to the output qubits.

\section{Circuit-noise simulation based on fault propagation}
\label{sec:standard_simulation}

Here, we review the standard circuit-noise simulation protocol based on the propagation of faults through the circuit.
The pseudo-code is provided in Algorithm~\ref{algo:standard_simulation}.
This is a detailed version of the procedure discussed in introduction. As explained earlier, we may have to repeat this process many times.

Following~\cite{delfosse2023spacetime}, denote by $\propagation F$ the cumulant of a fault operator $F$. It is the fault operator whose component $\propagation{F}_{\ell + 0.5} \in \Pauli_n$ after level $\ell$ is the result of all the faults occurring during the first $\ell$ levels of the circuit propagated through the first $\ell$ levels of unitary gates.
The cumulant can be computed by conjugating faults through unitary gates using the standard stabilizer simulation algorithm~\cite{gottesman1997stabilizer}.

\begin{algorithm}[ht]
\DontPrintSemicolon
\SetKwInOut{Input}{input}\SetKwInOut{Output}{output}
\Input{
A Clifford circuit $\circuit$, 
a noise model $\Prob_\circuit$,
a syndrome matrix $\mat_s$, 
a logical matrix $\mat_\ell$,
a decoder $D$,
an integer $n_{\sample}$.
}
\Output{A Monte-Carlo estimation of the failure rate of the circuit $\Prob_\circuit(D(s) \neq \bar f)$.}
\BlankLine
    Initialize $n_{\fail} = 0$. \;
    \For{$i=1,2,\dots n_{\sample}$}
    {
        Sample a fault operator $F$ according to the circuit-noise distribution $\Prob_\circuit$.\;

        Compute the cumulant $\propagation F$. \;
        
        Compute the effect $f = \eff_m(F)$ using $\propagation{F}$. \;
        
        Compute the syndrome $s$ using $s^T = \mat_s f^T$. \;
        
        Compute the logical flips $\bar f$ using $\bar f^T = \mat_\ell f^T$. \;
        
        If $D(s) \neq \bar f$, do $n_{\fail} \leftarrow n_{\fail} + 1$. 
    }
    \Return $\frac{n_{\fail}}{n_{\sample}}$.
\caption{Standard circuit-noise simulation}
\label{algo:standard_simulation}
\end{algorithm}

For simplicity, we separate the computation of the cumulant $\propagation F$ and the computation of its effect $f$ on measurement outcomes.
In practice, we do not need to store the whole cumulant in memory to obtain $f$.
It is enough to compute the levels of the cumulant $\propagation{F}_{\ell + 0.5}$ sequentially which reduces the memory cost of the simulation.
In this note, we ignore the memory cost because it is not the bottleneck of the simulation and it is not a clear differentiator between the algorithms discussed here.

Suppose that we have a fast decoder and consider the cost of the other steps of this simulation.
We provide the worst-case computational complexity of steps 3 to 7 of Algorithm~\ref{algo:standard_simulation} in two settings: (i) for a general circuit and (ii) for a sparse circuit made with unitary gates and measurement supported on a bounded number of qubits.
We compute the worst-case complexity as a function of the number of qubits $n$, the depth $\Delta$, the number of measurements $n_m$, the number of syndrome bits $n_s$ and the number of logical outcome bits $n_\ell$.
Many noise models produce low-weight fault operators $F$.
If this is the case, we can speed up some of the steps of the simulation.
Then, we provide the complexity as a function of the weight $|F|$ of the fault operator sampled.
Table~\ref{tab:complexity} summarizes our results.

{\it Generation of the matrices $\mat_s$ and $\mat_\ell$.}
In this work, we assume that the syndrome matrix $\mat_s$ and the logical matrix $\mat_\ell$ are given as an input with the circuit.
An algorithm generating a syndrome matrix, that is checks of the outcome code, is described in~\cite{delfosse2023spacetime}.
For general circuit, it runs in $O(n^4\Delta)$ bits operations. 
Its complexity is reduced to $O(n \Delta)$ for LDPC spacetime codes and $O(n)$ for periodic LDPC spacetime codes.
The logical matrix $\mat_\ell$ can be obtained by backpropagating the logical operators measured at the end of the circuit.

{\it Complexity of step 3 - Sampling $F$.}
The cost of sampling $F$ depends on the details of the circuit-noise distribution $\Prob_{\circuit}$.
For most popular noise models such as the phenomenological model or the circuit-level noise model of~\cite{dennis2002topological}, the complexity of generating a random fault operator $F$ is linear in the volume of the circuit, that is $O(n\Delta)$.

{\it Complexity of step 4 - Computation of $\propagation F$.} The standard approach to compute the cumulant $\propagation F$ of $F$ is through fault propagation.
Assume that a unitary gate $U$ acting on $w$ qubits is represented by a $2w \times 2w$ binary matrix that stores the conjugation $U P U^{-1}$ of the Pauli operators $P = X_q$ and $Z_q$ acting on a qubit $q$ of the support of $U$.
Conjugating a general Pauli fault $Q$ through this gate is equivalent to applying this matrix to the binary representation of the fault $Q$. This can be done in $O(w^2)$ operations.
As a result, the worst-case computational complexity of the computation of the cumulant with this approach is $O(n^2\Delta)$ bit operations for a general circuit (at most $O(n^2)$ per level).
For a sparse circuit, the complexity drops to $O(n\Delta)$ bit operations.
The computation of the cumulant of a low-weight fault operator with this method is not significantly faster because a single fault may rapidly spread to many qubits after propagating through a few levels of the circuit. Even though $F$ is sparse, it is generally not the case for its cumulant.
Better scaling is achieved in some cases such as circuits implemented exclusively with Pauli measurements, or fault-tolerant circuits designed to avoid spreading faults.

{\it Complexity of step 5 - Computation of $f$.}
Given $\propagation F$, the effect $f = \eff_m(F)$ can be obtained with a worst-case complexity of $O(n\Delta)$ bit operations for a general circuit. 
More precisely, the $j$th bit $f_j$ of $f$ is 1 iff $F$ induces a flip of the outcome of the $j$th measurement of the circuit.
Let $S_j$ be the measured operator and let $\ell_j$ be the level of this measurement.
Then, $f_j$ is obtained as the commutator of $\propagation{F}_{\ell_j-0.5}$ with the $j$th measured operator $S_j$ because the outcome of the measurement of $S_j$ is flipped iff the faults accumulated before this measurement ($\propagation{F}_{\ell_j-0.5}$) anti-commute with $S_j$.
This corresponds to Lemma~3 of~\cite{delfosse2023spacetime}, but we include it here because it is also used later on.
Recall that $\eta_{\ell_j - 0.5}(S_j)$ is the fault operator obtained by placing the operator $S_j$ right before level $\ell_j$.

\begin{lemma} [Lemma~3 of \cite{delfosse2023spacetime}]
    \label{lemma:effect_on_outcome}
Let $F$ be a fault operator.
The faults corresponding to $F$ induce a flip of the measurement of $S_j$ iff $[\propagation F, \eta_{\ell_j - 0.5}(S_j)] = 1$.
\end{lemma}
In other words, we have $f_j = [\propagation F, \eta_{\ell_j - 0.5}(S_j)]$.
Recall that following~\cite{delfosse2023spacetime}, we use the notation $[P, Q] = 0$ if $P$ and $Q$ commute and $1$ if they anti-commute.

The worst-case complexity of the computation of $f$ remains unchanged for a sparse circuit or for a low-weight fault operator $F$ because $\propagation F$ typically has large weight.

{\it Complexity of steps 6 and 7 - Computation of $s$ and $\bar f$.}
For a general circuit, the computation of the syndrome $s$ requires applying a $n_m \times n_s$ binary matrix to $f$ which can be done with a worst-case complexity of $O(n_m n_s)$.
Again, this subroutine is not significantly faster for sparse circuits or for low-weight fault operators.
Similarly, the computation of the logical flips $\bar f$ can be done with a complexity of $O(n_m n_\ell)$ bit operations in the worst case.
However, the number of logical outcome bits $n_\ell$ is often small.

Putting things together, we obtain a worst-case complexity in $O(n^2 \Delta^2)$ dominated by the syndrome computation. For some circuits, the syndrome and the logical flips can be computed more efficiently (in $O(n\Delta)$) and then the worst-case complexity of the simulation is in $O(n^2\Delta)$, or $O(n\Delta)$ for sparse circuits, dominated by the computation of the cumulant $\propagation F$ through fault propagation.

\section{ABC simulation: Circuit-noise simulation without fault propagation}
\label{sec:sim_wo_prop}

For some specific circuits, such as the standard syndrome extraction circuit for surface codes, we can bypass some of the steps of this simulation by directly computing the syndrome without fault propagation.
This is because the standard surface code circuit~\cite{tomita2014low} has a natural graph structure where each vertex corresponds to the measurement of an ancilla qubit in the circuit.
This is less obvious for other surface code circuits like the measurement-based circuits of~\cite{chao2020optimization} or~\cite{gidney2022pair} or for Floquet codes~\cite{hastings2021dynamically}.

In what follows, we propose a general algorithm to perform circuit-noise simulations without fault propagation; see Algorithm~\ref{algo:simulation_without_fault_propagation}.
This can be seen as a hypergraph generalization of the graph-based simulation technique for surface codes.
We also simplify the computation of the syndrome and the logical flips, resulting in a more favorable worst-case complexity for the simulation of arbitrary Clifford circuits.
Table~\ref{tab:complexity} compares the complexity of Algorithm~\ref{algo:standard_simulation} and Algorithm~\ref{algo:simulation_without_fault_propagation}.
At a high level, these algorithm are similar. One can think of Algorithm~\ref{algo:simulation_without_fault_propagation} as obtained by removing steps 4 and 5 from Algorithm~\ref{algo:standard_simulation}.
In Table~\ref{tab:complexity}, we use the step numbers of Algorithm~\ref{algo:standard_simulation} (step 4, 5, 6, 7) to compare these two strategies even though step 4 and 5 are not present in Algorithm~\ref{algo:simulation_without_fault_propagation}.

\begin{table}[ht]
{\small
    \centering
    \begin{tabular}{|c|c|c|c|c|c|c|}
    \hline
        Algorithm
        & Assumption
        & Precomp.
        & Comp. of $\propagation{F}$
        & Comp. of $f$
        & Comp. of $s$
        & Comp. of $\bar f$
        \\
        
        &
        &
        & (Step 4)
        & (Step 5)
        & (Step 6)
        & (Step 7)
        \\
    \hline
        Naive
        & General circuit
        & None
        & $O(n^2 \Delta)$
        & $O(n \Delta)$
        & $O(n_m n_s)$
        & $O(n_m n_\ell)$
        \\
    \hline
        Naive
        & Sparse circuit
        & None
        & $O(n \Delta)$
        & $O(n \Delta)$
        & $O(n_m n_s)$
        & $O(n_m n_\ell)$
        \\
    \hline
        ABC sim.
        & General circuit
        & $O((n_s + n_\ell)n\Delta)$
        & None
        & None
        & $O(n_m n_s)$
        & $O(n_m n_\ell)$
        \\
    \hline
        ABC sim.
        & LDPC ST code
        & $O(n_\ell n \Delta)$
        & None
        & None
        & $O(|F|)$
        & $O(|F| n_\ell)$
        \\
    \hline
        ABC sim.
        & LDPC ST code
        & $O(n)$
        & None
        & None
        & $O(|F|)$
        & $O(|F| n_\ell)$
        \\
        
        & + periodic
        &
        & 
        &
        &
        &
        \\
    \hline
    \end{tabular}
    \caption{Worst-case computational complexity of the main steps of the circuit-noise simulation for different classes of circuit with a naive simulation (Algorithm\ref{algo:standard_simulation}) and with ABC simulation (Algorithm~\ref{algo:simulation_without_fault_propagation}). We consider general Clifford circuits, sparse Clifford circuits made with circuit operations acting on a bounded number of qubits, Clifford circuits with an LDPC spacetime (ST) code and periodic circuits with an LDPC ST code.
    The worst-case complexity is computed as a function of the number of qubits $n$, the circuit depth $\Delta$, the number of measurements $n_m$, the number of syndrome bits $n_s$ and the number of logical outcome bits $n_\ell$. Our approach is favorable when the sample size is large or when the number of Pauli faults $|F|$ in the circuit is small.
    We do not include the cost of sampling (step 3) in this table because it is the same for all algorithms.
    }
    \label{tab:complexity}
}
\end{table}

\subsection{Case of general circuits}

In this section, we propose a circuit-noise simulation algorithm without fault-propagation for general Clifford circuits.
The procedure is described is Algorithm~\ref{algo:simulation_without_fault_propagation}.

The key ingredient to remove the fault propagation is the following result from~\cite{delfosse2023spacetime}. It relates the accumulator $F \mapsto \propagation F$ and the back-accumulator $F \mapsto \backpropagation F$. Recall that $\backpropagation F$ is the fault operator defined in the same way as $\propagation F$ but by propagating faults backward through the circuit.

\begin{prop} [Adjoint of the cumulant] [Proposition~3 of~\cite{delfosse2023spacetime}]
    \label{prop:key_prop_adjoint} 
For all fault operators $F, G$ of a circuit $\circuit$, we have
$$
[\propagation F, G] = [F, \backpropagation G] \cdot
$$
\end{prop}

The accumulator $F \mapsto \propagation{F}$ and the back-accumulator
can $F \mapsto \backpropagation{F}$ are linear operators acting on the space of Pauli operators on $n(\delta+1)$ qubits.
This Pauli group is isomorphic with $\Z_2^{2n(\delta+1)}$ (ignoring the global phase),
and it is equipped with the symplectic inner product defined by $[P, Q] = 0$ if $P$ and $Q$ commute and $1$ if they anti-commute.
Proposition~\ref{prop:key_prop_adjoint} states that the back-accumulator is the adjoint of the accumulator.
This is because of this key property for our simulation algorithm that we name it adjoint-based code (ABC) simulation.

For any vector $u \in \Z_2^{n_m}$, define the operator 
\begin{align} \label{eq:def_F(u)}
F(u) = \prod_{j = 1}^{n_m} \eta_{\ell_j - 0.5}(S_j^{u_j})
\end{align}
also used in~\cite{delfosse2023spacetime}.
Therein, $S_j$ is the $j$th measured operator and $\ell_j$ is the level of the circuit at which this operator is measured.
With this notation, Proposition~\ref{prop:key_prop_adjoint} leads to the following result.

\begin{cor}
    \label{cor:s_f_bits_evaluation}
Let $F$ be a fault operator with effect $f = \eff_m(F)$ on measurement outcomes.
If $u \in \Z_2^{n_m}$, then 
\begin{equation} \label{eq:bits_computation_wo_propagation}
    (f | u) = [F, \backpropagation{F(u)}]
\end{equation}
\end{cor}
Therein $(x | y) = \sum_i x_i y_j \pmod 2$ the standard binary inner product between two binary vectors.

\begin{proof}
By Lemma~\ref{lemma:effect_on_outcome}, we have $f_j = [\propagation F, \eta_{\ell_j - 0.5}(S_j)]$.
Using the standard properties of the commutator (see Section~3.3 of~\cite{delfosse2023spacetime}), we find
\begin{align}
    (u | f) 
    & = \sum_{j=1}^{n_m} u_j f_j \\
    & = \sum_{j=1}^{n_m} u_j [\propagation F, \eta_{\ell_j - 0.5}(S_j)] \\
    & = \sum_{j=1}^{n_m} [\propagation F, \eta_{\ell_j - 0.5}(S_j^{u_j})] \\
    & = [\propagation F, \prod_{j=1}^{n_m} \eta_{\ell_j - 0.5}(S_j^{u_j})] \\
    & = [\propagation F, F(u)]
\end{align}
and applying Proposition~\ref{prop:key_prop_adjoint}, we reach $[F, \backpropagation{F(u)}]$
\end{proof}

By definition, any bit $b$ of the syndrome $s$ or of the logical flips $\bar f$ can be written as $b = (u | f)$ for some vector $u \in \Z_2^{n_m}$ ($u$ is a row of $\mat_s$ or $\mat_\ell$).
As a result, one can directly compute any bit of $s$ or $\bar f$ without fault propagation and even without computing the effect $f$. Instead, we precompute $\backpropagation{F(u)}$ for each of the $n_s + n_\ell$ rows of the matrices $\mat_s$ and $\mat_\ell$.
The worst-case complexity of this precomputation grows as $O((n_s + n_\ell) n^2 \Delta)$ for a general Clifford circuit and is $O((n_s + n_\ell) n \Delta)$ for a sparse circuit.
Once the precomputation is done, each syndrome bit is obtained by computing a commutator with an operator $\backpropagation{F(u)}$ acting on at most $n(\Delta+1)$ qubits.

With this approach, the cost of the computation of $\propagation F$ and $f$ is removed and the worst-case complexity of computing $s$ and $f$ remains respectively $O(n_m n_s)$ and $O(n_m n_\ell)$ bit operations.

\begin{algorithm}[ht]
\DontPrintSemicolon
\SetKwInOut{Input}{input}\SetKwInOut{Output}{output}
\Input{
A Clifford circuit $\circuit$, 
a noise model $\Prob_\circuit$,
a syndrome matrix $\mat_s$, 
a logical matrix $\mat_\ell$,
a decoder $D$,
the precomputed operators $\backpropagation{F(u)}$ for each row $u$ of $\mat_s$ and $\mat_\ell$,
an integer $n_{\sample}$.
}
\Output{A Monte-Carlo estimation of the failure rate of the circuit $\Prob(D(s) \neq \bar f)$.}
\BlankLine
    Initialize $n_{\fail} = 0$. \;
    \For{$i=1,2,\dots n_{\sample}$}
    {
        Sample a fault operator $F$ according to the circuit-noise distribution $\Prob_\circuit$.\;

        Compute the syndrome $s$. 
        The $j$th bit of $s$ is $s_j = [F, \backpropagation{F(u)}]$ where $u$ is the $j$th row of the matrix $\mat_s$.\;
                
        Compute the logical flips $\bar f$.
        The $k$th bit of $\bar f$ is $f_k = [F, \backpropagation{F(u)}]$ where $u$ is the $k$th row of the matrix $\mat_\ell$.\;
        
        If $D(s) \neq \bar f$, do $n_{\fail} \leftarrow n_{\fail} + 1$. 
    }
    \Return $\frac{n_{\fail}}{n_{\sample}}$.
\caption{ABC simulation: Circuit-noise simulation without fault propagation}
\label{algo:simulation_without_fault_propagation}
\end{algorithm}

\subsection{Case of LDPC an spacetime code}

Assume now that the spacetime code of the circuit is LDPC. Recall that the spacetime code is defined by the stabilizer generators $\backpropagation{F(u)}$ used to compute the syndrome bits~\cite{delfosse2023spacetime}.
Because this code is LDPC, each qubit belongs to at most $O(1)$ of the operators $\backpropagation{F(u)}$ used to compute $s$.
Then, we can compute $s$ using the relation~\eqref{eq:bits_computation_wo_propagation} in $O(|F|)$ bit operations in the worst case.
The same argument does not apply to the computation of the bits of $\bar f$ because the corresponding back-accumulated operator can have large weight, however $n_\ell$ is often a small constant making the computation of $\bar f$ inexpensive.

If, in addition, the circuit we simulate is obtained by repeating a constant depth circuit periodically.
Then, the precomputation of the operators $\backpropagation{F(u)}$ used to compute $s$ and $\bar f$ can be done in $O(n)$ bit operations.

\section{Application to the simulation of large noisy Clifford circuit}
\label{sec:large_sim}

In this section we argue that, using ABC simulation, the simulation of a noisy Clifford circuit with large depth acting on many logical qubits is not significantly more expensive than the simulation of a single logical operation of this circuit.

To assess the performance of a large noisy Clifford circuit, it is common to simulate a small piece of this circuit.
For instance, to understand the performance of a quantum error correction code for building a quantum memory, we would like to run the quantum error scheme until a logical error occurs to estimate the lifetime of a quantum state.
Instead, we often simulate a single or a small number of logical cycles and we estimate the logical error rate per logical cycle~\cite{raussendorf2007fault, fowler2009high}. 
Other pieces of fault-tolerant quantum computing circuits have been simulated such as a lattice surgery operation on two logical qubits~\cite{vuillot2019code} and recently up to four logical qubits~\cite{bombin2023logical}.
This removes the need to propagate faults through a long circuit, making the simulation easier.
However, the results extrapolated from the performance of these small subcircuits of a larger circuit are not as accurate as a simulation of the whole lifetime of an encoded quantum state because
(i)~estimating the logical error rate of $T$ cycles by multiplying the logical error rate of a single logical cycle by $T$ is a rough approximation,
(ii)~the residual noise at the end of a logical cycle may affect the performance of the subsequent correction cycles and we cannot observe this phenomenon if we simulate a single logical cycle,
(iii)~the noise model may change and the noise rate may increase during the execution of a large circuit.

As an example, assume that each logical qubit is encoded in a patch of qubits for a code equipped with a fault-tolerant lattice surgery operation to perform logical Clifford gates and an efficient decoder (think of a surface code or a Floquet code patch for example).
Consider a circuit $\circuit$ starting the fault-tolerant preparation of all the logical qubits in the state $\ket{\bar 0}$, followed with $\Delta$ layers of lattice surgery operations\footnote{Suppose that most logical qubits are part of a lattice surgery operation and few logical qubits are idle.} and ending with the logical measurement of all the logical qubits.
This circuit produces a $N$-bit logical outcome.
The probability of an error on this logical outcome bit-string after decoding is some constant $\varepsilon \in [0, 1]$ that we want to estimate.

A standard way to estimate $\varepsilon$ is to compute the logical error rate $\varepsilon'$ of a single lattice surgery circuit $\circuit'$ acting on two logical qubits and to multiply $\varepsilon'$ by the number of lattice surgery operations.
Because there is about $\frac{N\Delta}{2}$ lattice surgery operations in the entire circuit $\circuit$, each of them has a logical error rate $\varepsilon'$ of the order of $\frac{2\varepsilon}{N\Delta}$.
To get a sufficiently small error-bar on our estimate of $\varepsilon'$, we must sample of the order of $\frac{N\Delta}{2\varepsilon}$ fault configurations in $C'$. 
Suppose that we use $\frac{15N\Delta}{\varepsilon}$ samples, so that we observe an average of $30$ logical errors.
Each fault configuration is obtained by generating a random single-qubit or two-qubit Pauli fault after each circuit operation\footnote{We assume that the circuit is made with single-qubit and two-qubit operations.} of $\circuit'$.
Overall, we need to produce $\frac{15 N \Delta |\circuit'|}{\varepsilon'}$ random Pauli faults (single-qubit or two-qubit) where $|\circuit'|$ is the number of single-qubit and two-qubit operations of the lattice surgery circuit.

Consider now the number of random Pauli faults needed to estimate $\varepsilon$ by simulating the entire circuit $\circuit$.
Because the noise rate $\varepsilon$ of the entire circuit is much higher than the noise rate of a single lattice surgery operation, we can achieve a similar error-bar as the previous strategy using only $\frac{30}{\varepsilon}$ samples. 
However, each sample requires generating a random Pauli fault for each operation of the entire circuit $\circuit$, that is about $\frac{N \Delta |\circuit'|}{2}$ operations.
Overall, the total number of random single-qubit or two-qubit Pauli faults needed to observe an average of 30 logical errors is again $\frac{15 N \Delta |\circuit'|}{\varepsilon}$, just like in the previous case.

The issue with the naive simulation method based on fault propagation is that one needs to propagate these faults through the entire circuit. Our simulation method without fault propagation removes this obstacle.
For this strategy to work, we need a decoder that can be executed efficiently for the whole circuit. This can be achieved with a sliding window decoder~\cite{dennis2002topological} which can be parallelized as proposed in~\cite{skoric2022parallel, tan2022scalable}.

The price to pay for the ABC simulation of the whole circuit using Algorithm~\ref{algo:simulation_without_fault_propagation} is the precomputation of the backpropagated operators $\backpropagation{F(u)}$ corresponding to the syndrome bits and the logical outcome bits.
As discussed previously (see Table~\ref{tab:complexity}), this cost is not significant in many cases because the operators $\backpropagation{F(u)}$ associated with syndrome bits are often sparse and there are only $N$ operators $\backpropagation{F(u)}$ associated with logical bits in $\circuit$.
Moreover, we can use the fact that the circuit $\circuit$ is made with the same subcircuits repeated many times (fault-tolerant preparation, lattice surgery and logical measurements) to speed up the backpropagation of the $F(u)$.

\section{Conclusion}

We proposed an ABC simulation algorithm for noisy Clifford circuits that removes the need for fault propagation and opens the way to the simulation of large noisy Clifford circuits.
In particular, our approach is ideal for direct simulation of long sequences of fault-tolerant logical operations on many logical qubits, which stands in contrast to rough extrapolation based on composition of small subcircuits.
It could, for instance, provide accurate estimates the performance of large circuits based on lattice surgery with surface codes or Floquet codes.
Prime candidates include the plethora of Floquet codes recently introduced~\cite{aasen2022adiabatic, davydova2023floquet, kesselring2022anyon, bombin2023unifying, townsend2023floquetifying, dua2023engineering, zhang2022x, ellison2023floquet, davydova2023quantum}.

ABC simulation is compatible and may be combined with variance reduction techniques such as ~\cite{bravyi2013simulation,aliferis2007subsystem,iyer2022diagnostics,iyer2018small}.
In the low noise-rate regime, this may speedup fault sampling and further reduce the number of required samples overall.


\end{document}